\documentstyle[preprint,prl,aps]{revtex}
\begin{document}
\draft
\title{Local electronic structure of a single magnetic 
impurity in a superconductor}
\author{Michael E. Flatt\'{e}}
\address{Department of Physics and Astronomy, 
University of Iowa, Iowa City, Iowa 52242}
\author{Jeff M. Byers} 
\address{Naval Research Laboratory, Washington D.C. 
20375 } 
\maketitle
\begin{abstract}
The electronic structure near a single classical magnetic impurity in
a superconductor is determined using a fully self-consistent
Koster-Slater algorithm.  Localized excited states are found within the
energy gap which are half electron and half hole. 
Within a jellium model we find the new result that the
spatial structure of the positive-frequency (electron-like)
spectral weight (or local
density of states), can differ strongly from that
of the negative frequency (hole-like) spectral weight.  
The effect of the impurity on the continuum states above the
energy gap is calculated with good spectral resolution for
the first time. 
This is also the first three-dimensional self-consistent calculation 
for a strong magnetic impurity potential.
\end{abstract} 
\vfill\eject
\narrowtext

Magnetic impurities have
a dramatic effect on  superconductivity. Most work on
the experimental and  theoretical effects of magnetic
impurities have focused on bulk thermodynamic quantities, such as the
reduction  of the superconducting transition temperature $T_c$ with
increasing magnetic impurity concentration\cite{Mathiass}.
Theoretical approaches to treat the thermodynamic effects of impurities
include Born scattering (Abrikosov-Gor'kov theory)\cite{AG} and
approximate
solutions to all orders in the impurity potential for 
classical\cite{Shiba,Rusinov,Sakurai} 
and quantum\cite{Kondo} spins. A key issue
addressed in Ref.~3 was the 
evolution of localized excited states
around impurities into an impurity band, eventually leading to gapless
superconductivity at high enough impurity concentrations.
The effects of impurities on bulk properties have also been
treated within a strong-coupling formalism, but not
self-consistently or beyond the Born approximation (e.g. 
Ref.~7).
Concern about bulk properties in the above and related work did not
extend to properties very near the impurity. 

Among the first properties calculated in the vicinity of an
impurity were the structures of screening clouds around a
charged impurity\cite{Prange,Hurault} and a magnetic
impurity\cite{Hurault,Anderson}
in a superconductor (characterized by exponentially-decaying
Friedel-like oscillations). The oscillation of the order
parameter around a magnetic impurity was first evaluated without
self-consistency\cite{Tsuzuki,Kummel}.
A self-consistent calculation
of the order parameter at the impurity and very far away 
for {\it weak} impurity potentials was done by
Schlottmann\cite{Schlottmann}.
Interest in local properties near impurities has been
revived by advances in scanning
tunneling microscopy (STM) near impurities embedded in a
metallic medium\cite{Crommie1}.

Motivated by the possibility
of measuring the local electronic structure near an
impurity in a superconductor with an STM, the differential
conductivity through an STM tip was calculated near
impurities for
superconductors with isotropic and anisotropic gaps within the Born
approximation\cite{BFS}. 
The differential conductivity 
measured at a point ${\bf x}$ and voltage $V$ and
temperature $T$ can be related
to the local density of states (LDOS) at the tip location as follows:
\begin{equation}
{dI({\bf x},V,T)\over dV} = {1\over N_o} \int_{-\infty}^\infty
d\omega {\partial n(\omega)\over \partial\omega}
\left({{\rm Im}G({\bf x},{\bf x};\omega=eV)\over \pi}\right).
\end{equation}
Here $e$ is the charge of the electron, $n(\omega)$ is
the Fermi occupation function at temperature $T$, 
$N_o$ is the density of states at the Fermi level, and
$dI/dV$ is in units of the normal-metal's differential
conductivity.
The local density of states is the imaginary part of the
retarded Green's function fully dressed by the interaction of the
electronic system with the impurity.
Within the Born approximation, then,  the differential
conductivity can be expressed in terms of the retarded homogeneous Green's
functions of the superconductor, $g({\bf x},{\bf x'};\omega)$
and $f({\bf x},{\bf x'};\omega)$ (where $f$ is the anomalous
Green's function), in the
following way: 
\begin{eqnarray}
{dI({\bf x},V)\over dV} \propto &&{\rm Im}\Bigg[g({\bf x},{\bf
x};eV)\nonumber\\ && +  g({\bf x},{\bf
0};eV)V_{0}g({\bf 0},{\bf x};eV) -  f({\bf x},{\bf
0};eV)V_{0}f({\bf 0},{\bf x};eV)\Bigg]\label{Born}
\end{eqnarray}
where the impurity is at the location ${\bf 0}$ in the solid and the
non-magnetic impurity potential is $V_0\delta({\bf x})$.
For a BCS
superconductor with an isotropic order parameter in a parabolic band,
for $\omega$ much smaller than the Fermi energy,
\begin{eqnarray}
g({\bf x},{\bf x'};\omega) =&&-{\pi N_o\over k_Fr} {\rm
e}^{-\sqrt{\Delta^2-\omega^2}k_Fr/\pi\xi\Delta}\left(\cos k_Fr + {\omega\over
\sqrt{\Delta^2-\omega^2}}\sin k_Fr\right)\nonumber\\
f({\bf x},{\bf x'};\omega) =&&-{\pi N_o\Delta\over
k_Fr\sqrt{\Delta^2-\omega^2}}{\rm e}
^{-\sqrt{\Delta^2-\omega^2}k_Fr/\pi\xi\Delta}\sin k_Fr
\label{bareg}\end{eqnarray} 
where $r = |{\bf x}-{\bf x'}|$, 
$\Delta$ is the order parameter, and $\xi=\hbar
k_F/\pi\Delta$ is the
coherence length. These expressions are valid for $\omega$ above and
below $\Delta$ so long as the imaginary parts of both $f$ and $g$
are multiplied by ${\rm sgn} \omega$.

The Born approximation calculation would not yield any localized
states
around the impurity, and does not consider the effect of the
change in electronic structure on the local order
parameter $\Delta({\bf x})$. Recent preliminary STM results indicate
certain features of the local density of states 
near a magnetic impurity\cite{Yazdani}:
(1) discrete states are evident within the energy gap and 
the LDOS associated with them is 
asymmetric with voltage around $V=0$, and (2) 
the LDOS becomes indistinguishable from the bulk
density of states within a distance greater than the Fermi wavelength
but much less than the coherence length. A calculation beyond the
Born approximation where the order parameter is self-consistently
determined and the continuum spectrum is found would be useful for
explaining these results.

We will present a fully self-consistent calculation of the local
electronic structure near a magnetic impurity which is based on a
Koster-Slater-like Green's function technique rather than the
Bogoliubov-de-Gennes (BdG) equations\cite{deGennes}. The BdG
equations are  Schr\"odinger-like equations for the electron
and  hole components of the quasiparticle wavefunctions. 
The localized 
states of a vortex core in a superconductor and  other properties
revealed by the LDOS were calculated self-consistently using the BdG
equations\cite{Shore,Gygi}. These results were
compared with a measurement of a single vortex by an 
STM on superconducting  ${\rm NbSe_2}$\cite{Hess}. The BdG equations were
successfully used  again to explain the STM observations of the vortex
lattice\cite{lattice}.

Since the 
original application of the Green's function algorithm to localized vibrational
modes\cite{phonons}, it has been applied to
numerous problems including conduction electrons in 
metals\cite{Koster} and impurity states in magnets\cite{Wolfram},
but not to superconductors.
To place this algorithm in context we will contrast it with the BdG
equations. 
To find the quasiparticle wavefunctions, typically 
the inhomogeneity is placed in a sphere
of radius $R$ with appropriate boundary conditions. 
The value of $R$ is determined by 
the spectral resolution necessary for accurately 
determining the order parameter self-consistently
and the spectral width of features
measurable by (for example) the STM.  The
typical complications resulting from approximating an infinite 
system by a finite-size system apply, such as 
discrete states above the energy gap and the 
heavy investment of computer time required for 
large values of $R$.

The Green's function method works within a 
sphere whose radius is determined by the range of the 
inhomogeneous potential.
In essence we shall invert the Dyson 
equation in real space. The Dyson equation for an 
inhomogeneity in a superconductor, the Gor'kov
equations\cite{AGD}, can be written 
as: \begin{equation}
\int d{\bf x''} \left[ {\bf \delta}({\bf x}-{\bf x''}) - 
{\bf g}_\sigma({\bf x},{\bf x''}, \omega){\bf V}({\bf 
x''})\right] {\bf G}_\sigma({\bf x''},{\bf x};\omega) = {\bf 
g}_\sigma({\bf x},{\bf x'};\omega)\label{KS}
\end{equation}
where 
\begin{equation}
{\bf g}_\sigma({\bf x},{\bf x'};\omega) = \left(\begin{array}{cc}g_\sigma({\bf
x},{\bf x'};\omega)&f_\sigma({\bf x},{\bf x'};\omega)\\
f_\sigma^*({\bf x},{\bf x'};-\omega)& -g_\sigma^*({\bf x},{\bf x'};-
\omega)\end{array}\right), \end{equation}
the same relationship exists among ${\bf G}_\sigma$, 
$G_\sigma$ and $F_\sigma$, and
\begin{equation}
{\bf V}({\bf x''}) = \left( 
\begin{array}{cc}
\sigma V_S({\bf x''})+ 
V_{0}({\bf 
x''})&\delta\Delta({\bf x''})\\ 
\delta\Delta({\bf x''})& \sigma V_S({\bf 
x''})- V_{0}({\bf x''})\end{array}
\right).\label{potn}
\end{equation}
$\delta\Delta({\bf x}) = 
\Delta({\bf x})-\Delta_o$, where $\Delta({\bf x})$ is the
position-dependent order parameter and $\Delta_o$ is the
order parameter of the homogeneous superconductor.
$V_S({\bf x})$ is a localized, spin-dependent
potential, such as one originating from a classical spin.
$V_{0}({\bf x})$ is a localized non-magnetic
potential. The quantization direction of the superconductor's
spins ($\sigma=\pm 1/2)$ is chosen
parallel to the direction of the classical spin.
For the purposes of heuristic arguments in this Letter we
will assume ferromagnetic coupling between impurity spin and
electron spin ($V_S<0$). Antiferromagnetic coupling only
produces the trivial change that the
antiparallel spin quasiparticle is attracted to the
classical spin.
$g_\sigma$ is given by Eqs.~(\ref{bareg}) with $N_o$
labelling the density of states for each spin. $G_\sigma \ne
G_{-\sigma}$ due to the differences in the potential in Eq.~(\ref{potn}).
The self-consistency equation is
\begin{equation}
\delta\Delta({\bf x}) = \gamma\sum_\sigma\int_{-\infty}^\infty d\omega n(\omega)
{\rm Im}F_\sigma({\bf x},{\bf x};\omega) - \Delta_o,\label{KSsc}
\end{equation}
where $\gamma$ is the pairing strength.

One strength of this formalism is its reliance on the 
short-range nature of the inhomogeneous potential. 
Solution of Eq.~(\ref{KS}) requires inverting the frequency-dependent 
real-space matrix 
${\bf M}(\omega) = {\bf \delta}({\bf x}-{\bf x'}) -
{\bf g}_\sigma({\bf x},{\bf  x'};\omega) {\bf V}({\bf x'})$.
For ${\bf x'}$'s where
${\bf V}({\bf x'})$ is negligible, that portion of ${\bf M}(\omega)$
is triangular and trivially invertible.  
Hence the 
numerical complexity of inverting 
${\bf M}(\omega)$
is  governed by the range of
${\bf V}({\bf x})$.  
The radius of the sphere the system is solved in
is governed by the range of the longest-range 
potential, which in this paper will be 
$\delta\Delta({\bf x})$. 

As a special case of Eq. (\ref{KS}), the
energies of the localized states within the gap correspond to those
$\omega>0$ where 
${\rm det}{\bf M}(\omega)=0$.
We model the impurity with a Gaussian $V_S({\bf x})$ of
range $k_F^{-1}$.
Figure 1 shows the dependence of the localized state energies for the
first two angular momentum channels on the strength of the magnetic
potential, 
$v_s=\pi N_o| \sigma \int d{\bf x}V_S({\bf x})|$.
For $V_{0}=0$, the Eqs.~(\ref{KS})-(\ref{KSsc}) are 
unchanged by the
transformation $\omega\rightarrow -\omega$ and $s \rightarrow -s$
so the poles of $G_\sigma(\omega)$ must come in symmetric pairs ($\sigma
= \pm 1/2$) around
$\omega=0$\cite{nonmag}. 
The quasiparticle state for small $v_s$ corresponding to
these poles consists of an electron ($\omega>0$) in the spin
band parallel to the classical spin, which we will label up
($\uparrow$), and a hole ($\omega<0$) in the other spin band
($\downarrow$). A hole in the down band has spin up and is
attracted to the classical spin. Spin is a good quantum
number for the quasiparticle, which is spin up. The
spatially-integrated spectral weight is one-half each for
the electron and hole components (this is true for all
localized quasiparticle states in this model).

For small $v_s$ an already existing
quasiparticle can be bound by the classical spin.
This allows for local
pair-breaking excitations of energy less than $2\Delta_o$.
As the potential strength increases, the excitation energy of each
angular momentum state $\ell$ decreases, at some critical value ($v_{s\ell}^*$)
vanishes, and then begins to increase.  
For $v_s>v_s^*$ the {\it ground} state contains a spin-up
quasiparticle bound to the classical spin\cite{Sakurai}.  
The low-energy excitation now corresponds to removing 
that quasiparticle, which is a spin-down excitation.
This qualitative behavior of the excitation energies can
be extracted from the Shiba model\cite{Shiba}, where the
magnetic potential is modeled by a
delta function at the impurity site. In Ref.~3 zeros of
$M(\omega)$ are
found, neglecting the component of ${\rm Re}{g}({\bf x},{\bf
x'};\omega)$ which is symmetric for $\omega\rightarrow
-\omega$. These poles are shown in Figure 1.

Even in the normal state the 
spectral weight of the up band has a peak 
at the origin, while that of the down band is pushed away
from the origin.  In the superconductor this asymmetry is
evident in the quasiparticle spectral weights.
Figure 2 shows the calculated angular momentum 
$\ell = 0$ spectral weights in the up and down bands
for two values of $v_s$, indicating that the
asymmetry of the spatial structure of the spectral weights
becomes more pronounced with increasing $v_s$. Also shown
are $\ell = 1$ spectral weights for $v_s=0.875$.  The
asymmetric localized-state spectral weights produce 
differential conductivities which are asymmetric as well\cite{surfcav}.
Figure 3 shows differential conductivities for $v_s=1.75$ at three
locations --- right at the impurity, $k_F^{-1}$ away, and
$10k_F^{-1}$ away. By $10k_F^{-1}$ the spectrum has
recovered to the homogeneous spectrum. The differential
conductivity farther away can be easily recovered by
constructing the self-consistent ${\bf T}$-matrix 
${\bf V}({\bf x'})\delta({\bf x'}-{\bf x''}) + {\bf
V}({\bf x'}){\bf G}({\bf x'},{\bf x''};\omega){\bf
V}({\bf x''})$ directly from the dressed Green's functions.
The spatial dependence of the differential conductivity
far away from the impurity
is qualitatively identical to the Born result\cite{BFS}.

The asymmetric behavior of the electron-like and hole-like spectral weights
due to the difference between the up band states and down
band states
does not emerge from the model of
Ref.~3, which implies identical spatial dependences
of the spectral weights
for the up and down bands. 
As pointed out by Koster and
Slater\cite{Koster}, proper treatment of the symmetric real
part of ${
g({\bf 0},{\bf 0};\omega)}$ is 
essential for obtaining the local electronic properties around an
impurity. 
The proper approximation for the real symmetric 
part of the Green's function is to average it over a small
volume given by the range of the potential modeled by the
delta function. We therefore introduce a new parameter into
the Shiba model,
$\alpha$, which is the $\omega$-symmetric part of  
${\rm Re}\langle g({\bf 0},{\bf 0};\omega=0)\rangle/\pi N_o$.
The brackets indicate that $g({\bf 0},{\bf x};\omega=0)$ 
has been averaged over a
small volume.
For $\alpha\ne 0$ the ratio of the
spectral weight of the up band at the impurity, $A_\uparrow({\bf
0})$, to the spectral weight of the down band at the impurity,
$A_\downarrow({\bf 0})$, is
\begin{equation}
{A_\uparrow({\bf 0})\over A_\downarrow({\bf 0}) } 
= {1+2\alpha v_s+(1+\alpha^2)v_s^2\over
1-2\alpha v_s+(1+\alpha^2)v_s^2}.\label{specrat}
\end{equation}
The introduction of $\alpha$ does not change the localized state
energies qualitatively.

We now discuss the structure of $\Delta({\bf x})$. 
Figure 4 shows 
$\delta\Delta({\bf x})$'s for two values of 
$v_s$.  While oscillating
with wavelength $\sim \pi k_F^{-1}$, $\delta\Delta({\bf x})$
falls off to a negligible potential within $10k_F^{-
1}$. This justifies performing the numerical calculation in
a sphere of that radius. 
A typical radial grid of 100 points provides a 
numerically robust solution. The self-consistent solution
also depends on the value of $N_o\Delta_o/k_F^3$, which for niobium is
$3.6\times 10^{-5}$\cite{Nb_DOS}.

As shown in Figure 4, for large values of $v_s$, $\Delta({\bf
x}={\bf 0})<0$. Sign changes in $\Delta$, as seen in pair
tunneling, have been suggested for magnetic impurities in
the barriers of Josephson junctions\cite{junctions}.
Our sign change in $\Delta({\bf 0})$ occurs (at $T=0$) 
precisely at $v_{s0}^*$. 
The symmetry of 
Eqs.~(\ref{KS})-(\ref{KSsc}) under $\omega\rightarrow
-\omega$ and $s\rightarrow -s$ implies that 
${\rm Im}F_\uparrow ({\bf r},{\bf r}, \omega) = -
{\rm Im}F_\downarrow ({\bf r},{\bf r}, -\omega)$. As the
pole in the spin-up band goes from electron-like ($\omega>0$) to
hole-like ($\omega<0$) and the pole in the spin-down band goes from 
hole-like to electron-like the contribution to $\Delta({\bf
0})$ changes sign abruptly. $\Delta({\bf 0})$
as a function of $v_s$ is shown in the insert of Figure 4. At $T>0$ the
transition would be smoothed somewhat.

The behavior of $\Delta({\bf 0})$ as a function of $v_s$ comes
from the introduction at $v_{s0}^*$ of a quasiparticle into the ground
state of the system. 
The spin up quasiparticle localized near
the impurity in the ground state suppresses the local order parameter.
Exciting the low-energy state (removing the spin-up
quasiparticle) causes a
pair to enter the condensate.  
We find that exciting the low-energy
state for $v_s>v_{s0}^*$ {\it increases} $\Delta({\bf 0})$,
whereas excitation of quasiparticles typically reduces
$\Delta({\bf x})$ (which is the case for $v_s<v_{s0}^*$.
Also, exciting the low-energy state {\it reduces} the induced spin
of the superconductor at the impurity.
 We note that this picture
should have implications for the theory of gapless
superconductivity, which is now based on the formation of
impurity bands through hybridization of single-quasiparticle
excited states around impurities, without consideration of
the coherence of excitations with the condensate.

We wish to acknowledge useful conversations with A.V.
Balatsky, B.A. Jones, M. Salkola, and A. Yazdani.
M.E.F. wishes to acknowledge the Office of Naval Research's
Grant No. N00014-96-1-1012. J.M.B. wishes 
to acknowledge an N.R.C. Fellowship.

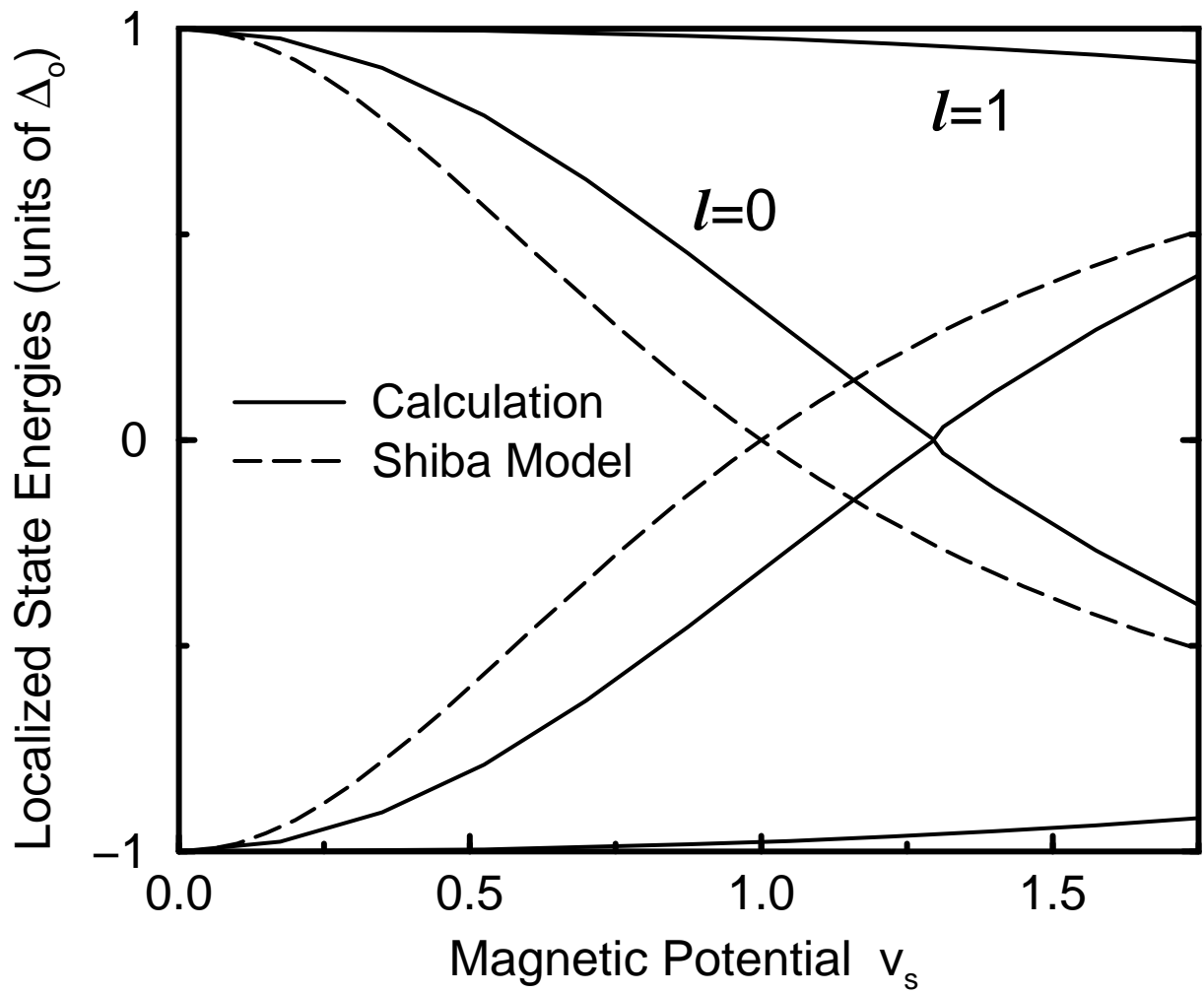
\begin{figure}
\caption{Solid lines indicate the
self-consistently calculated localized excited
state energies for angular momentum channels $\ell =0$ and
$1$.  The analytic model of Ref.~3 is shown by the dashed
lines.  At a critical value of
$v_s=v_{s0}^*$, the character of the $\ell = 0$ state changes
from spin up to spin down.
The kink evident in the solid line is real, and is due
to the discontinuous change (at $T=0$) 
in $\Delta({\bf x})$ at
$v_{s0}^*$.}
\end{figure}
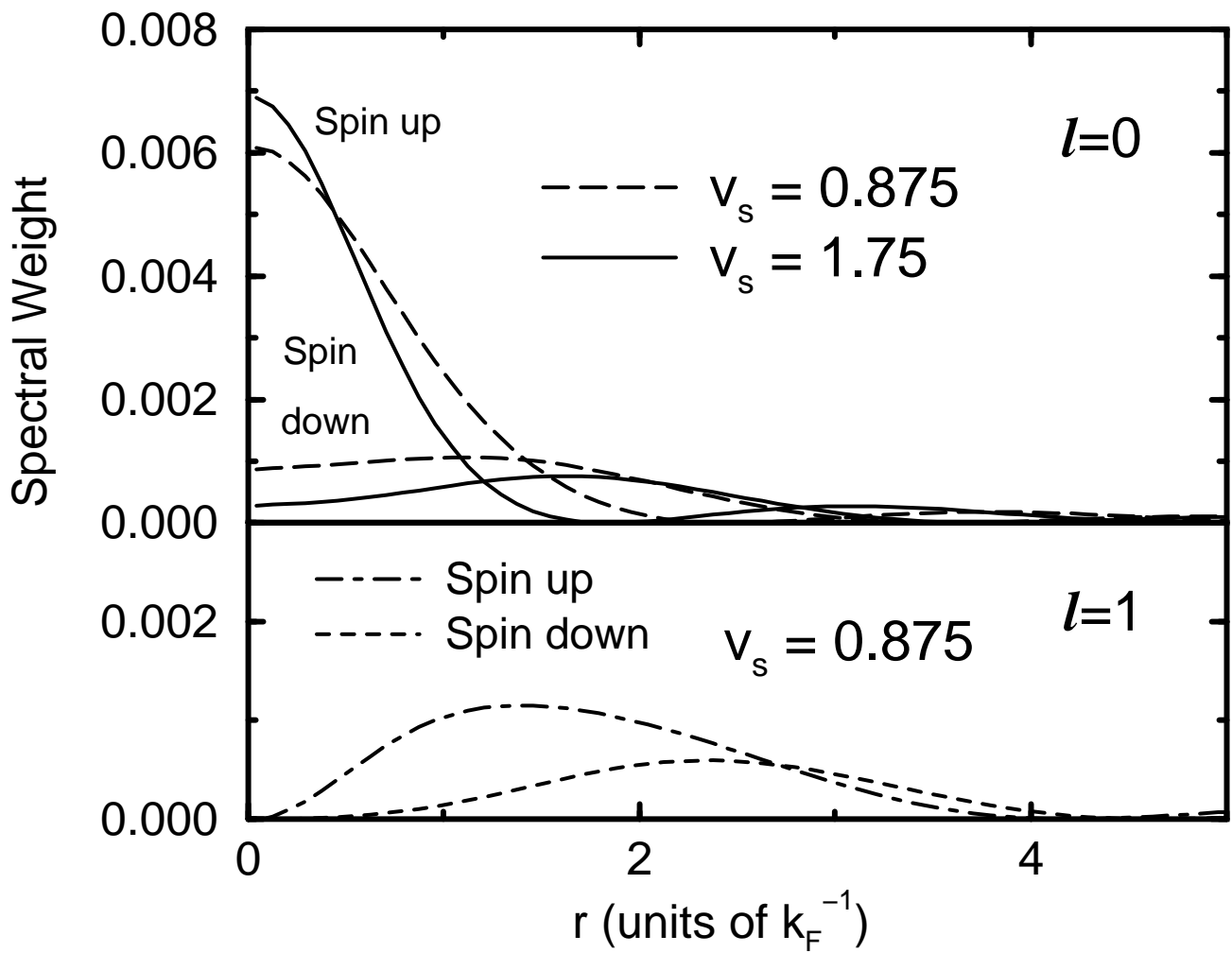
\begin{figure}
\caption{Spectral weights for the $\ell = 0$ state's
up-band component (electron-like for $v_s<v_s^*$) and down-band 
component (hole-like for $v_s<v_s^*$)
as a function of position for two values of the
magnetic potential. Also shown are the spectral weights for
the up and down
bands for $\ell=1$ for $v_s = 0.875$.  }
\end{figure}
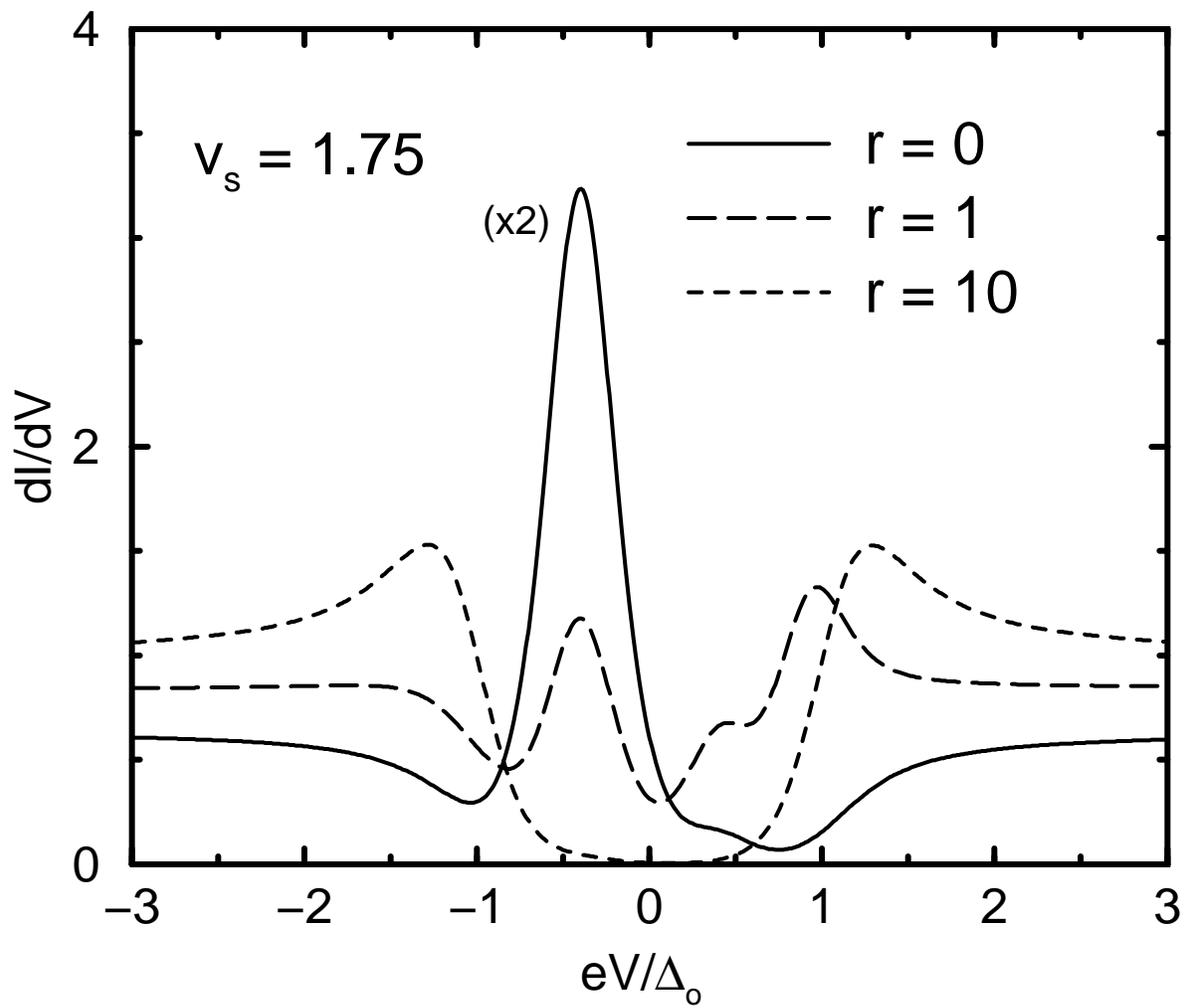
\begin{figure}
\caption{Differential conductivity ($dI/dV$), relative to the normal
metal at three distances (in units of $k_F^{-1}$)
from the impurity for $v_s=1.75$,
showing the evolution of the spectrum from one dominated by
the localized states near $r=0$ to one dominated by the bulk
spectrum at $r=10$. The curve at $r=0$ has been shrunk by a
factor of two so that it can appear on the same scale
as the other two curves.  The presence of the large peak near
the origin on the negative-frequency side of the spectrum
indicates that the impurity is strong ($v_s>v_{s0}^*$).
$dI/dV$ has been evaluated from Eq.~(1) with
$\beta=7.5/\Delta_o$, which for niobium corresponds to
$T\sim 2$K.}
\end{figure}
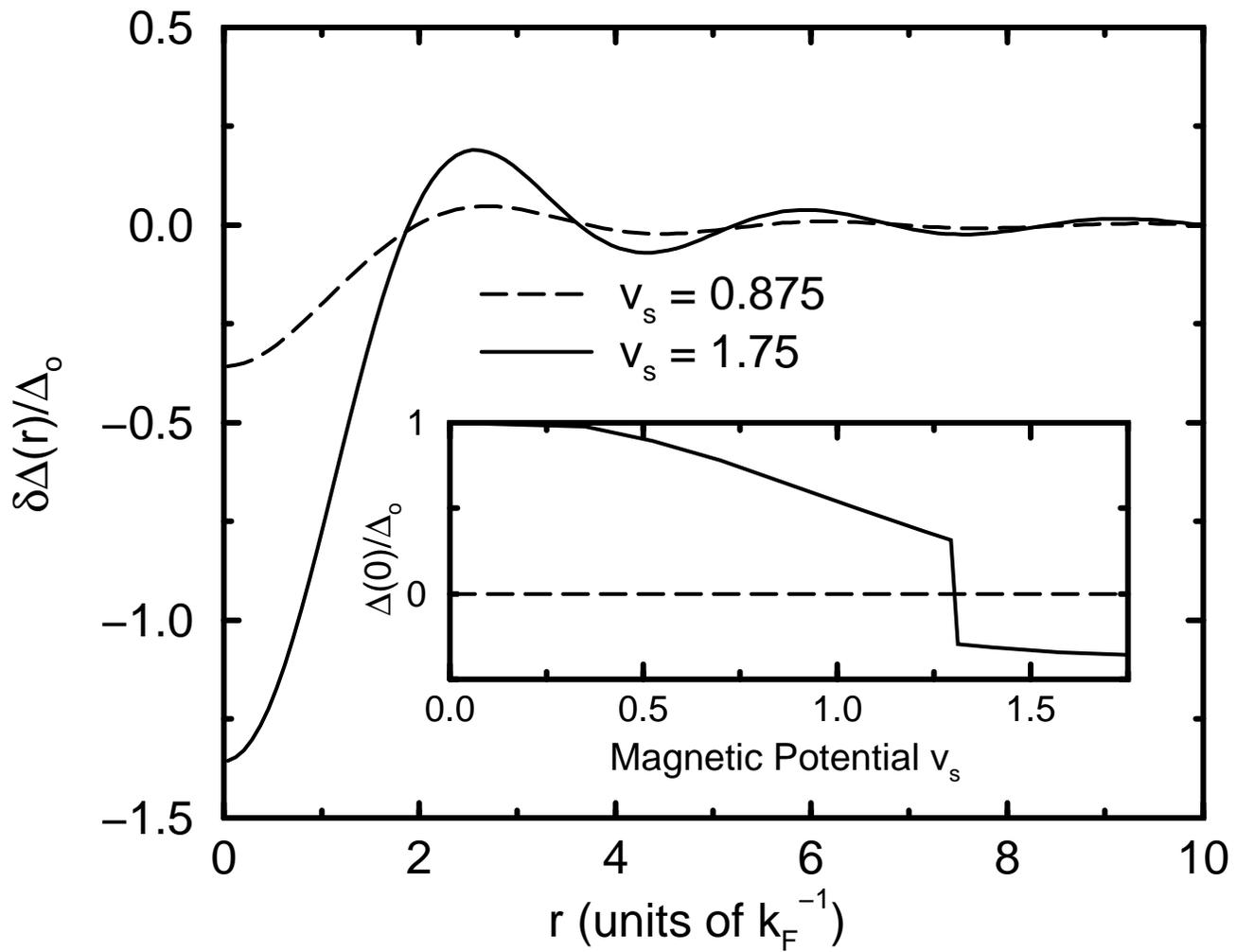
\begin{figure}
\caption{Change in the local order parameter, $\delta\Delta({\bf x})$,
for two values of the magnetic potential. The change becomes
negligible beyond $10k_F^{-1}$. (Insert) Order parameter at
the impurity, $\Delta({\bf 0})$, as a function of $v_s$,
indicating a discontinuous change at $v_{s0}^*$.  }
\end{figure}


\begin{references}
\bibitem{Mathiass} B. T. Matthias, H. Suhl, and E. Corenzwit, Phys. Rev. 
Lett. {\bf 1}, 92 (1958),
		C. Herring, Physica {\bf 24}, S 184 (1958) and
		H. Suhl and B. T. Matthias, Phys. Rev. {\bf 114}, 977 
(1959).
\bibitem{AG} A. A. Abrikosov and L. P. Gor'kov, Zh. Eksp. Teor. Fiz. {\bf 
39}, 1781 (1962) (Soviet Phys. JETP {\bf 12}, 1243 
(1961)).
\bibitem{Shiba} H. Shiba, Prog. Theor. Phys. {\bf 40}, 435 (1968).
\bibitem{Rusinov} A. I. Rusinov, Soviet Phys. JETP Letters
{\bf 9}, 85 (1969).
\bibitem{Sakurai} A. Sakurai, Prog. Theor. Phys. {\bf 44},
1472 (1970).
\bibitem{Kondo} O. Sakai, {\it et al.}, J. Phys. Soc. Japan
{\bf 62}, 318 (1993).
\bibitem{Carbotte} F. Marsiglio, J.P. Carbotte, A. Puchkov,
and T. Timusk, Phys. Rev. B {\bf 53}, 9433 (1996).

\bibitem{Prange} R. Prange, Phys. Rev. {\bf 129}, 2495
(1963).

\bibitem{Hurault} J. P. Hurault, Journal de Physique {\bf
26}, 252 (1965).

\bibitem{Anderson} P.W. Anderson and H. Suhl, Phys. Rev.
{\bf 116}, 898 (1959).

\bibitem{Tsuzuki} T. Tsuzuki and T. Tsuneto, Prog. Theor.
Phys. {\bf 37}, 1 (1967).

\bibitem{Kummel} R. Kummel, Phys. Rev B {\bf 6}, 2617 (1972).

\bibitem{Schlottmann} P. Schlottmann, Phys. Rev. B {\bf 13},
1 (1976).

\bibitem{Crommie1} M.F. Crommie, C.P. Lutz, and D.M. Eigler, Nature 
(London), {\bf 363}, 524 (1993).  
Y. Hasegawa and P. Avouris, Phys. Rev. Lett. {\bf 71}, 1071 (1993).

\bibitem{BFS} J.M. Byers, M.E. Flatt\'e, D.J. Scalapino, Phys. Rev.
Lett. {\bf 71}, 3363 (1993).

\bibitem{Yazdani} A. Yazdani, C. P. Lutz and D. E. Eigler, unpublished.

\bibitem{deGennes} See, e. g., P. G. de Gennes, {\it Superconductivity 
of Metals and Alloys} (Addison-Wesley, Reading, MA 1989). 

\bibitem{Shore} J. D. Shore, M.Huang, A. T. Dorsey and J. P. 
Sethna, Phys. Rev. Lett. {\bf 62}, 3089 (1989).

\bibitem{Gygi} F. Gygi and M. Schl\"{u}ter, Phys. Rev. B {\bf 41}, 822 
(1990) and F. Gygi and M. Schl\"{u}ter, Phys. Rev. B {\bf 43}, 7609 
(1991).

\bibitem{Hess} H. F. Hess, R. B. Robinson, R. C. Dynes, J. M. Valles, Jr., 
and J. V. Waszicak, Phys. Rev. Lett. {\bf 62}, 214 (1990); J. Vac. 
Sci. Tchnol. A {\bf 8}, 450 (1990).

\bibitem{lattice} H. Hess, R. B. Robinson, and J. V. Waszczak,
Phys. Rev. Lett. {\bf 64},2711 (1990) and F. Gygi and 
M. Schl\"{u}ter, Phys. Rev. Lett. {\bf 65}, 1820 (1990).

\bibitem{phonons} E. W. Montroll and R.B. Potts, Phys. Rev.
{\bf 100}, 525 (1955).

\bibitem{Koster} G.F. Koster and J.C. Slater, Phys. Rev.
{\bf 95}, 1167 (1954).  G.F. Koster and J.C. Slater, Phys. Rev.
{\bf 96}, 1208 (1954).

\bibitem{Wolfram} T. Wolfram and J. Callaway, Phys. Rev.
{\bf 130}, 2207 (1963).

\bibitem{AGD} See A.A. Abrikosov, L.P. Gor'kov, and I.E. Dzyaloshinski, 
{\it Methods of Quantum Field Theory in Statistical Physics}
(Dover, New York, 1963).

\bibitem{nonmag} Even when $V_0\ne 0$ the discrete-state
energies come in symmetric pairs.

\bibitem{surfcav} We expect the local electronic structure
around an impurity on a surface, which is measurable by STM,
to be qualitatively the same as the structure around an
impurity embedded in the bulk.

\bibitem{Nb_DOS} ($k_F = 1.18\AA^{-1}$, $\xi = 380\AA$) 
K. -H. Hellwege and J. L. Olsen, eds., {\it
Landolt-B{\"o}rnstein} {\bf 13c}
(Springer, Berlin, 1984).

\bibitem{junctions} See references in M. Sigrist and T.M.
Rice, Rev. Mod. Phys. {\bf 67}, 503 (1995).

\end{references}
\end{document}